\documentclass[chicago]{emulateapj} 

\usepackage{amssymb}
\usepackage{color}

\shorttitle{Implosion of coronal loops during a solar flare}
\shortauthors{Sim\~oes, Fletcher, Hudson and Russell}

\slugcomment{Accepted by ApJ on 26 September 2013}
\begin{document}

\title{Implosion of coronal loops during the impulsive phase of a solar flare}

\author{P.~J.~A.~Sim\~oes\altaffilmark{1}, L.~Fletcher\altaffilmark{1}, H.~S.~Hudson\altaffilmark{1,2}, A.~J.~B.~Russell\altaffilmark{1,3}} 
\email{email: paulo.simoes@glasgow.ac.uk}
\altaffiltext{1}{SUPA, School of Physics and Astronomy, University of Glasgow, G12 8QQ, UK}
\altaffiltext{2}{SSL, UC Berkeley, CA 94720, USA}
\altaffiltext{3}{Present address: Division of Mathematics, University of Dundee, Dundee, DD1 4HN, UK}

\begin{abstract} {We study the relationship between implosive motions
    in a solar flare, and the energy redistribution in the form of
    oscillatory structures and particle acceleration. The flare
    SOL2012-03-09T03:53 (M6.4) shows clear evidence for an
    irreversible (stepwise) coronal implosion. Extreme-ultraviolet
    (EUV) images show at least four groups of coronal loops at
    different heights overlying the flaring core undergoing fast
    contraction during the impulsive phase of the flare.  These
    contractions start around a minute after the flare onset, and the
    rate of contraction is closely associated with the intensity of
    the hard X-ray (HXR) and microwave emissions.  They also seem to
    have a close relationship with the dimming associated with the
    formation of the Coronal Mass Ejection (CME) and a global EUV
    wave. Several studies now have detected contracting motions
    in the corona during solar flares that can be interpreted as the
    implosion necessary to release energy.  Our results confirm this,
    and tighten the association with the flare impulsive phase. We add
    to the phenomenology by noting the presence of oscillatory
    variations revealed by GOES soft X-rays (SXR) and
    spatially-integrated EUV emission at 94 and 335 {\AA}. We identify
    pulsations of $\approx 60$ seconds in SXR and EUV data, which we
    interpret as persistent, semi-regular compressions of the
    flaring core region which modulate the plasma temperature and
    emission measure. The loop oscillations, observed over a large
    region, also allow us to provide rough estimates of the energy
    temporarily stored in the eigenmodes of the active-region
    structure as it approaches its new equilibrium.}
\end{abstract}


\section{Introduction}
\label{sec:intro}
Understanding the interplay of the magnetic field evolution, both in
the corona and in the photosphere, and the time evolution of
electromagnetic emissions produced during a flare can provide
information about the nature of the energy-release process and its
location.  The energy for flares and CMEs comes from the coronal
magnetic field, stored in regions displaying magnetic stress, and
we can observe the evolution of these regions via their emissions.
The extraction of the magnetic energy $\int B^2/8\pi \mathrm{d}V$ must
correspond in a general sense to a reduction in the magnetic pressure,
and consequently the magnetic structure must contract so as to achieve
a new equilibrium position.  This scenario was proposed by
\citet{2000ApJ...531L..75H}, suggesting that such implosions of the
coronal magnetic field could be detected with extreme-ultraviolet
(EUV) observations, such as TRACE.  We present here an excellent
example of this in the flare SOL2012-03-09 (M6.4) observed by the Solar
Dynamics Observatory \citep{2012SoPh..275....3P}, showing inward
motions and oscillations of loops both during and following the
implosion.  The event has well-resolved hard X-ray (HXR) sources that
locate the main regions of energy output, and microwave emission
which can sensitively identify the onset of particle acceleration.

Observations of such magnetic contractions in the energy-rich
impulsive phase are still rare.  We distinguish these from the
``shrinkage'' observed during the gradual phase of a flare
\citep{1996ApJ...459..330F} and attributed to large-scale magnetic
reconnection in a current sheet created by a flare eruption.  We
review some examples here, highlighting what is known about the
relative timing of flare, contraction and CME.
\citet{2009ApJ...696..121L} presented observations of contractions of
three clusters of EUV coronal loops overlying the flaring region
during the early impulsive phase of SOL2005-07-30.  The contraction
speed measured from {\em TRACE} 171~{\AA} data was slow: about 4--7
km~s$^{-1}$ over about 10 minutes during the impulsive phase.  This was
accompanied by converging motion of the HXR footpoints. The
  downward motion of coronal HXR sources in the impulsive phase of
  solar flares, first reported by \citet{2003ApJ...596L.251S}, can also
  be associated with converging footpoint motions
  \citep[e.g.][]{2004ApJ...611L..53L}, making it likely that the HXR
  source's downward motion is part of the implosion phenomenon.
  Slow
coronal loop contraction was also observed during the impulsive phase
of SOL2003-10-24, with average speed of 6-10~km~s$^{-1}$ \citep{2009ApJ...706.1438J}.
\citet{2006A&A...446..675V} observed downward motions of hard X-ray
sources in SOL2003-11-03, and \citet{2007ApJ...660..893J} describe such motions 
as contraction resulting from the reduction of sheared field. 
\citet{2009ApJ...703L..23L} report {\em TRACE} 195~{\AA} loop
contraction sustained for $\sim$12 minutes in flare SOL2001-06-15.  
In this case the loop contraction occurred after the impulsive phase and
was well associated in time with a filament eruption.  The authors
also argue that the collapsing loops formed a shrinking trap which
then accelerated hot electrons to non-thermal energies, producing a
new coronal HXR source a few minutes after the onset of the
contraction.  \citet{2010ApJ...714L..41L} observed a fast (100~km~s$^{-1}$)
loop contraction with {\em TRACE} 171~{\AA} delayed by about 200
seconds from the impulsive phase of SOL2005-09-08, but preceded by a
much slower contraction phase.  \citet{2012ApJ...757..150L} reported
contracting and erupting (expanding) components in five flares
originating from sigmoidal active regions, from GOES classes B to X,
all observed with SDO/AIA.  Contracting coronal loops overlying
the ends of the sigmoid were observed in cooler AIA channels (i.e. 171
and 193~{\AA}) while expanding bubbles were observed in warm/hot AIA
channels (i.e. 94, 211, and 335~{\AA}).  Contraction speeds varied
from 10~km~s$^{-1}$ to over 200~km~s$^{-1}$, with higher speeds corresponding to
larger events. The eruption always preceded the contraction though the
time delay was smaller for more energetic events, in which eruption
was also concurrent with the HXR impulsive phase.
 
The well-observed flare SOL2011-02-15T01:50 (X2.2) has provided some
of the clearest evidence for the implosion.
\citet{2012ApJ...748...77S} reported EUV loops collapsing and
oscillating during this flare, and made the
important identification of this evolution with permanent changes in
the photospheric magnetic field.  \citet{2012ApJ...749...85G}
identified three distinct phases of coronal loop dynamics in this
event: a slow rise phase when the loops expanded prior to the flare; a
rapid collapse phase of loop contraction, with the lower loops
collapsing before the higher ones; and an oscillation phase when the
loops displayed oscillations interpreted as global kink modes.
This followed the collapse phase, and the period of oscillation
increased with the height of the loop.

There have been many observations of flare-related loop oscillations
or pulsations, observed in direct imaging and also in flare
lightcurves at different wavelengths. Rather than summarise the literature here (reviews have been given by \citet{2005LRSP....2....3N,2005RSPTA.363.2743D,2007SoPh..246....3B,2012RSPTA.370.3193D})
we highlight a few salient properties,
focusing on events which appear to have an abrupt onset clearly
related to a flare.  From EUV imaging, oscillations appear to be
triggered in a number of ways, for example the destabilisation and
ejection of a filament which runs into a distant system of loops and
excites them \citep[examples given in][]{2002SoPh..206...69S}, the
displacement of overlying loops as a global shock wave (as represented by
a type II burst) impinges on them \citep{2004ApJ...614L..85H}, the
oscillation of loops relaxing post-reconnection
\citep{2012A&A...545A.129W} or in association with a magnetic implosion
\citep{2010ApJ...714L..41L,2012ApJ...748...77S}.  Our
loop oscillations fall in this last category.

The majority of the imaging observations reported have been in the
EUV, sampling loops with an estimated temperature of 1-1.5~MK,
e.g. from the Transition Region and Coronal Explorer \citep[TRACE;
][]{1999SoPh..187..229H} but UV spectroscopic observations by
\citet{2002ApJ...574L.101W} and \citet{2003A&A...406.1105W}, using the
Solar Ultraviolet Measurement of Emitted Radiation instrument
\citep[SUMER; ][]{1995SoPh..162..189W} have detected oscillatory
Doppler-shift signatures in lines from ions at formation temperatures
in excess of 6~MK, identified as standing slow-mode oscillation.
High-temperature oscillating loops (9-11~MK) were detected using
SDO/AIA by \citet{2012A&A...545A.129W}.  The authors described the
excitation as a release or relaxation of a loop following large-scale
magnetic reconnection, explaining their high temperatures as
consistent with a post-flare coronal plasma.

Non-imaging observations, particularly of flare HXR and microwave
lightcurves, show oscillatory structure often termed `pulsations' and
related to particle dynamics \citep[e.~g.][]{1970SoPh...13..420C}.
Since HXR emission is only weakly directional for steep spectra, pulsations in the HXR
flux are probably due to variations either in the electron
acceleration rate, or possibly in the precipitation rate in a
periodically compressing and expanding loop.  Microwave emission from
non-thermal electrons, on the other hand, is strongly directional, so
microwave pulsations could also be due to a global kink mode, which
would generally change the line-of-sight angle, or even due a global
sausage mode which would change the field strength in the emitting
region \citep[e.~g.][]{2009SSRv..149..119N}.  Pulsations in line and
continuum emissions in soft X-rays have not been frequently remarked
upon \citep[but see][]{2006ApJ...639..484M, 2012ApJ...749L..16D}; we
discuss here their presence in the GOES data for the event we study.
Variations at these wavelengths require changes in plasma temperature,
density or volume, so they may also require a compressional mode, or a
rapid sequence of heating/cooling events at different locations.

\section{Observational data and analysis}
\label{sec:data}

The flare SOL2012-03-09 (M6.4) occurred around 03:40UT, in active
region NOAA 11429 (N17W12), and it was observed by many observatories: in
microwaves by the {\em Nobeyama Radio Heliograph} \citep[\em{NoRH},
][]{1994IEEEP..82..705N} and {\em Nobeyama Radio Polarimeter} \citep[\em{NoRP}, ][]{1985PASJ...37..163N},
 in HXR by {\em Reuven Ramaty
  High Energy Solar Spectroscopic Imager} \citep[\em{RHESSI},][]{2002SoPh..210....3L} and {\em Fermi Gamma-ray Burst
  Monitor} \citep[\em{GBM}, ][]{2009ApJ...702..791M}, in soft X-rays (SXR) by
{\em Geostationary Operational Environmental Satellite (GOES)}. High
resolution images in EUV taken by the {\em Atmospheric Imaging
  Assembly (AIA)}, and line-of-sight magnetograms from {\em
  Helioseismic and Magnetic Imager (HMI)}, both on board of SDO were
also used in our analysis. As described below, these observations
produced beautiful examples of large-scale loop contraction, wave
excitation, dimming, and complex oscillations of loop structures.
\citet{2012A&A...544L..17H} described this event as a white-light
flare, underscoring the intensity of its energy release.

\subsection{Pre-flare configuration of AR 11429}
\begin{figure*}
\plotone{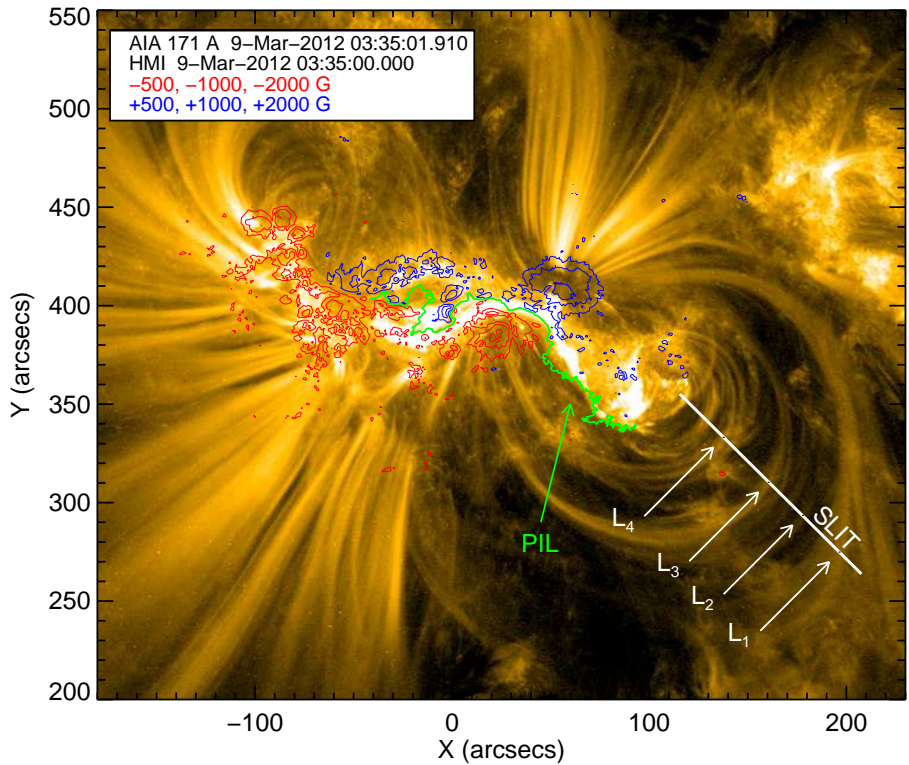}
\caption{The configuration of active region NOAA 11429 at 03:35~UT, before
  the flaring activity, as seen by the SDO/AIA 171~{\AA} filter,
  overlaid by contours of the photospheric line-of-sight magnetic
  field by SDO/HMI at the same time (the contour levels are indicated
  on the figure), with a well-defind polarity inversion line
  (PIL). The white line indicates the slit taken to create the
  time-position diagram (see Fig.~\ref{fig:wiggles}), and the arrows
  point to the four collapsing bundles of loops ($L_1$ to $L_4$). See the electronic edition
of the Journal for a color version of this figure. This figure is also
available as an mpeg animation in the electronic
edition of the Astrophysical Journal.
\label{fig:pre1}}
\end{figure*}

The magnetic configuration of the active region at the photospheric
level was classified as $\beta \gamma \delta$.  Fig.~\ref{fig:pre1}
shows the line-of-sight HMI magnetograms and the polarity inversion
line (PIL) in the context of an AIA image at 171~{\AA}.  Some of the
stronger magnetic kernels are connected by large-scale coronal loops,
with their apexes projected as much as 100~Mm to the southwest
of the region's core.  Their orientation is not clear, but they
appear to arch around rather than over the active-region core, and
thus may be very inclined with respect with the local vertical; other
strongly magnetic regions appear to link to still larger-scale
structures.  
The southwest loop structures, as labeled in
Fig.~\ref{fig:pre1}, exhibit the contraction and oscillations discussed below.  
Prior to any flaring activity we can identify a dark filament in
171~{\AA} following the magnetic polarity inversion line (PIL), which
seems to connect the eastward positive polarity and the westward
negative polarity.  Between 03:00$-$03:25~UT bright kernels appear
along the filament.  At 03:25~UT, brightenings associated with an
M1.8 flare, followed by an M2.1 flare, appear at the east and west
roots of the filament and along the filament itself, extending along the
PIL. During this early interval prior to the main M6.4 event, the dark
filament evolves to become a bright and highly sheared sigmoid
structure.  The temperature derived from GOES soft X-rays shows that
the two smaller flares peak at $18-19$~MK and $\approx 17$~MK,
cooling to $\approx 15$~MK before the main event; note that this
earlier activity does not directly result in the dynamics discussed
below.
\subsection{Coronal implosion}
\begin{figure*}
\plottwo{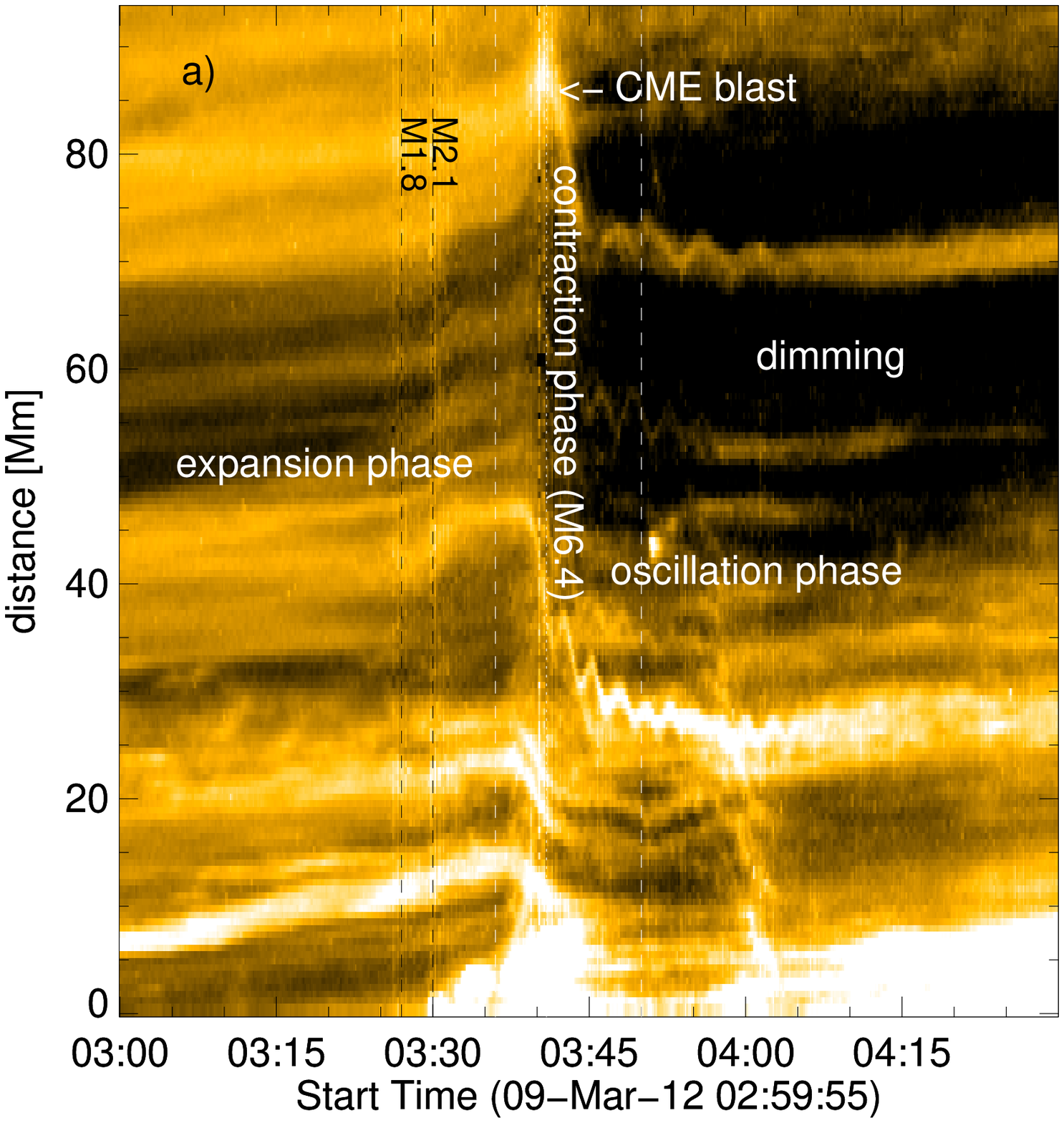}{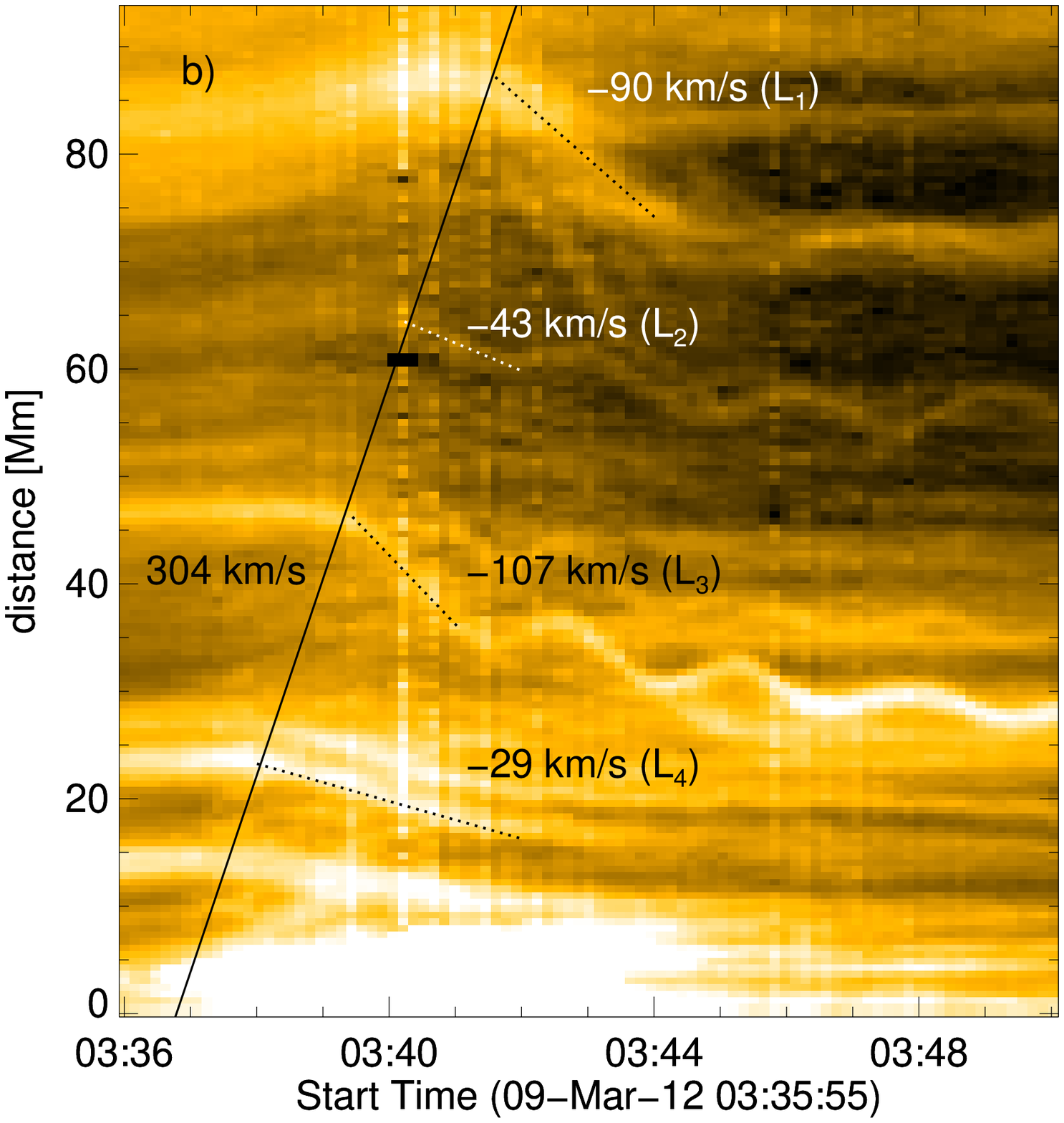}
\caption{a) Time-position diagrams for the slit in
  Fig.~\ref{fig:pre1}, showing the pre-flare configuration, the
  quick expansion phase after the two preceding flares, the main
  collapse and oscillations phase. The coronal dimming after the
  passage of a EUV wave is clearly visible, as it is the expansion of
  the outer loop, which could be related to the CME launch. b) Same as
  the left-hand frame, zoomed in during the impulsive phase (white
  vertical dashed lines in the left frame), showing the average speed
  of the initial implosion of four identified loops and the delay of
  the implosion onset with height. See the electronic edition
of the Journal for a color version of this figure. \label{fig:wiggles}}
\end{figure*}

In Fig.~\ref{fig:pre1} we show the position of a visible set of loops
before the main impulsive phase of the flare. The group of large-scale
loops, as seen at 171~{\AA} (also seen in 131, 193 and 211 {\AA}) and
rooted in strong-field parts of the flaring region, rapidly contract
during the impulsive phase of the flare.  We can identify at least
four loops (or narrow bundles of loops) marked $L_1$ to $L_4$, from
greater to smaller distance from the AR core).  A slit 153 pixels long
($\approx 92''$) and 3 pixels wide ($\approx 2''$) was taken across
the group of contracting coronal loops (see Fig.~\ref{fig:pre1}) and
the time evolution of the intensity in AIA 171~{\AA} images at the four
locations along this slit is plotted in Fig.~\ref{fig:wiggles}a.  This
time-position diagram shows the development of the EUV features: a
pre-flare quiet period revealing a slow expansion of the EUV features,
followed by a faster expansion after the two smaller flares (M1.8 at
03:27~UT and M2.1 at 03:34~UT), and a main contraction phase which
includes the beginning of the oscillation phase.  The three outer
loops show evident oscillations even as they contract, with distinct
periods and phases (Fig.~\ref{fig:wiggles}b).  It is not possible to
tell from the data if the inner set $L_4$ oscillates after the
contraction. 
%
The onset and duration of the intense EUV emission below 10~Mm in
Fig.~\ref{fig:wiggles}b agrees in time with the flare impulsive phase,
which continues (as shown by HXR and microwave emission, as discussed
later) throughout the main contraction of loops at all heights.  Each
loop set seems to contract at a different projected speed of
approximately $90$, $43$, $107$, and $29$ km~s$^{-1}$ (for loops $L_1$ to
$L_4$) (see Fig.~\ref{fig:wiggles}b).  The $L_1$ motion appears not
only to be related to the general contraction, but also to the CME
launch: A rapid expansion of the outer loops $L_1$ prior to the time
marked ``contraction phase'' in Fig.~\ref{fig:wiggles}a is evident and
thereafter seemingly try to return to their original
positions. Moreover, the coronal dimming coinciding with the passage
of the EUV wave and CME launch is also clear.

The loops at different heights do not start to contract at
the same time, showing evident delays of 60--80 seconds.  The
difference between initial and final projected heights of the loop
sets are 13, 12, 20 and 6 Mm, from $L_1$ to $L_4$ respectively.
Higher/longer loops show longer periods of oscillation, and loops that
undergo greater displacements also contract faster, although a
relationship between projected height and contracting speeds cannot be
identified from these data.

\begin{figure}
\plotone{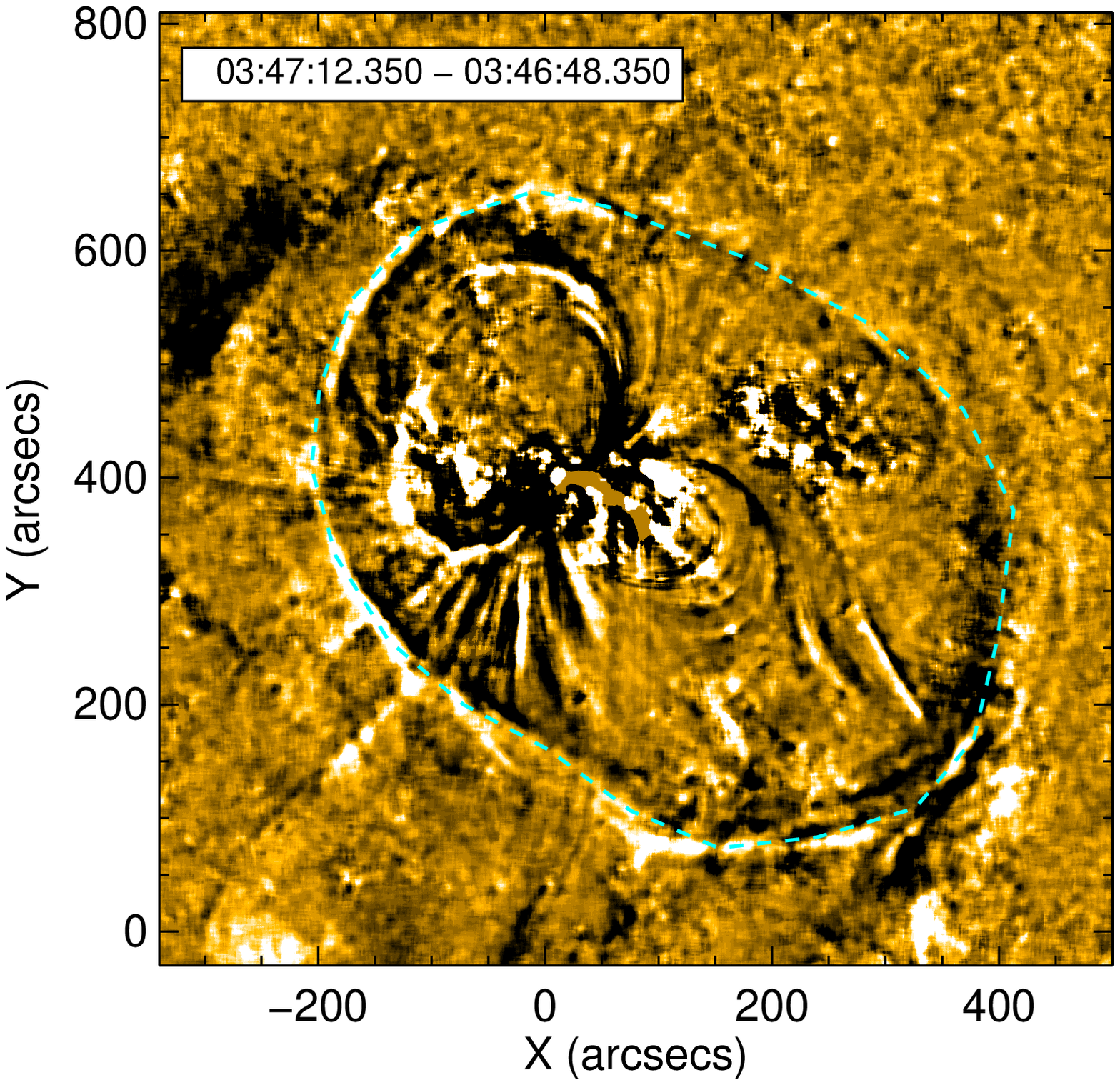}
\caption{Difference image from the AIA 171~{\AA} band, with the
  expanding wave front designated by the dashed line. See the electronic edition
of the Journal for a color version of this figure. This figure is also
available as an mpeg animation in the electronic
edition of the Astrophysical Journal.
\label{fig:wave_snapshot}}
\end{figure}

The movie representations of the AIA images show a global wave,
roughly concentric with the core of the active region, which we
illustrate with the snapshot in Fig.~\ref{fig:wave_snapshot} at 171
{\AA} (also seen in 193 and 211 {\AA}).  The presence of a large-scale
shock disturbance, a reasonable interpretation of the EUV images, is
consistent with the reported occurrence of a meter-wave type~II burst
in this event.

\begin{figure}
\plotone{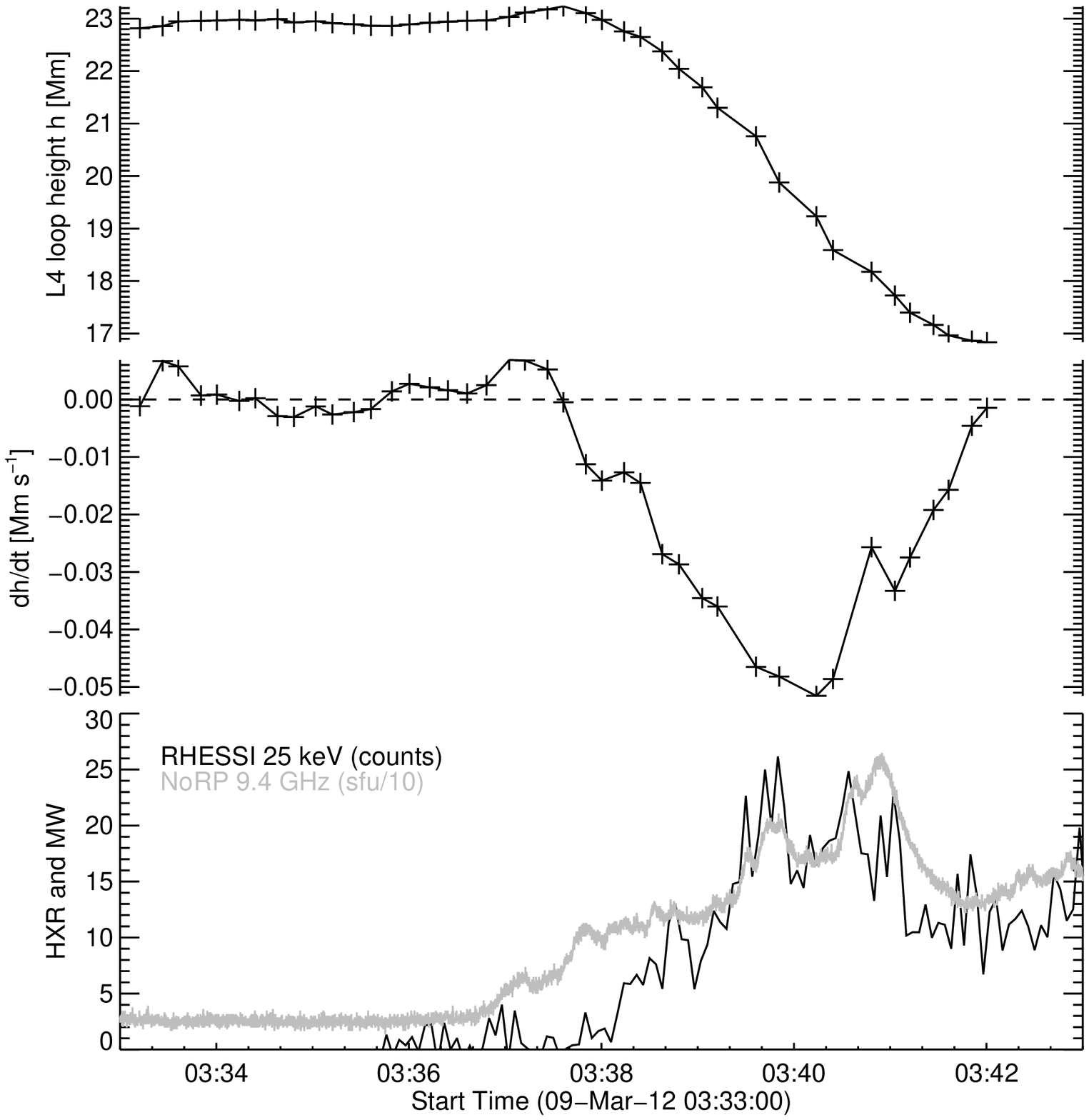}
\caption{Timing of the contraction of coronal loops and flare emission
  (top), and the rate of change $dh/dt$ (middle), compared to the
    impulsive phase observed in HXRs from RHESSI, and 9.4~GHz
    microwaves from NoRP (bottom).
\label{fig:timing}}
\end{figure}

The onset of the contraction for loops L1 to L4 can be reasonably well
identified.  As shown in Fig.~\ref{fig:timing}, it is clear that the
contraction of the loops starts after the flare onset, which can be
identified by the almost simultaneous increase in the flare HXR and
microwave lightcurves at 03:36:45~UT while the L4 loop starts to
collapse close to 03:38~UT. If we track back the contraction onset
times to zero distance using a straight line, as shown in the right
hand panel of Fig.~\ref{fig:wiggles} this gives an intercept also at
around 03:36:45~UT.  Note that this substantially precedes the time
indicated as ``contraction phase'' in the Figure. Also note that the
rate of contraction $dh/dt$ (middle panel of Fig.~\ref{fig:timing}) is
well associated with the non-thermal emission, reaching its maximum at
the peak of the flare.

If we assume a common triggering mechanism for the implosion of the
entire magnetic structure containing the four loops, the information
travels outwards at a projected speed of $\approx 300~$km~s$^{-1}$ from the
core of the active region.  This speed, derived from the straight line
connecting the onset of contraction of each one of the four loops, is
only an order-of-magnitude estimate, due to projection effects of
the inclined EUV loops.  Nevertheless, it is clear that the
contraction of the magnetic structure starts after the flare onset,
and successively later at greater projected distances.

The outer EUV loops ($L_1$, $L_2$, and $L_3$) continue to contract rapidly 
(and start to oscillate) after the HXR peak (around 03:40~UT). 
This is all consistent with the association of energy release, into visible
forms, with an implosion of the magnetic structure that had contained that energy.

\subsection{UV/EUV ribbons and coronal loops}

The ribbon/footpoint structures represent the site of the dominant
impulsive-phase energy release, and so their association with the
development of the coronal implosion provides a guide to the
mechanisms of energy transport and dissipation.

\begin{figure*}
\plotone{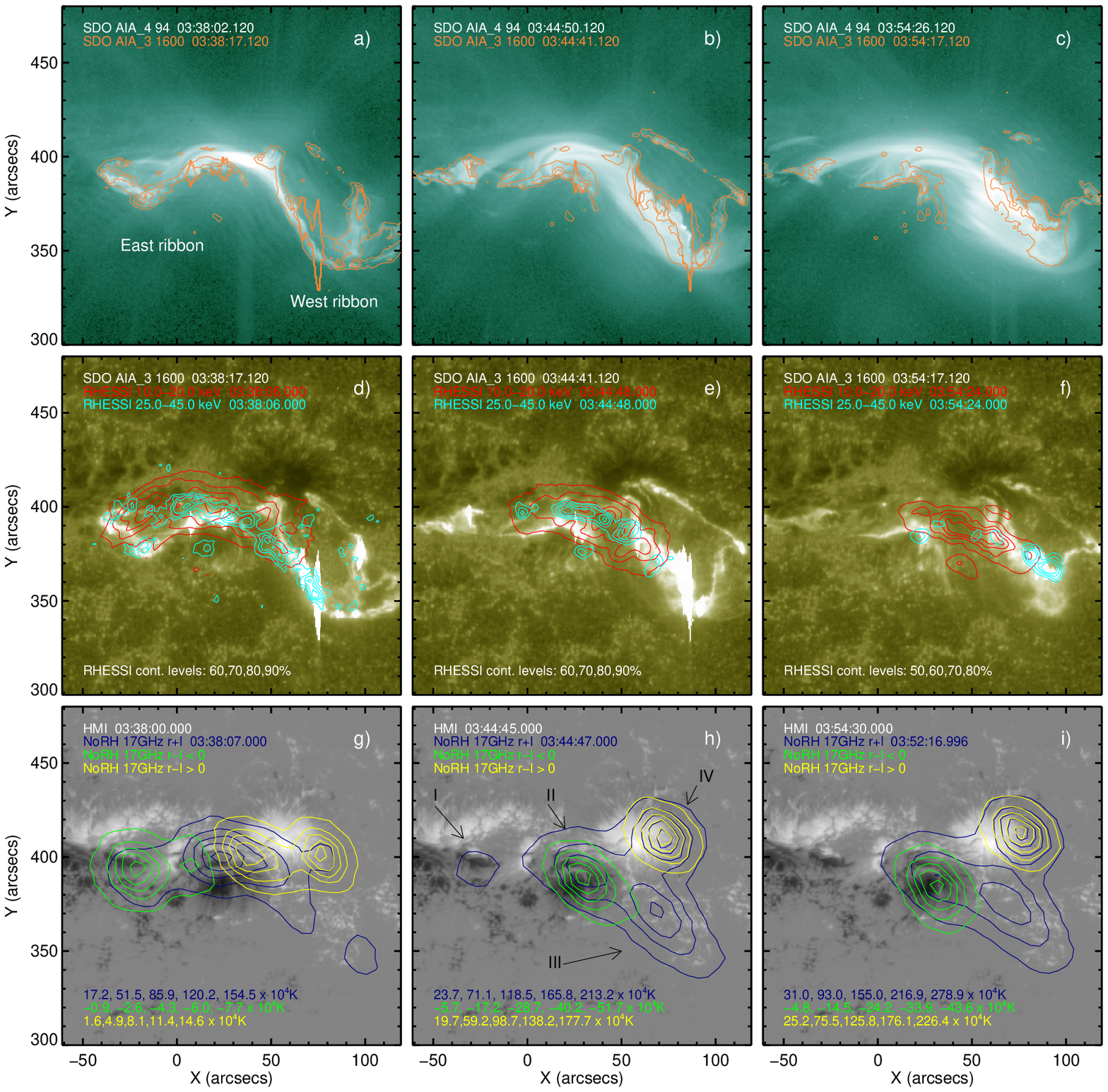}
 \caption{Spatial morphology of the flare at three time intervals:
   rise phase (first column), after the peak (second column) and in the
   decay phase of the second burst (third column). (a--c) AIA 94~{\AA}
    images overlaid with 1600~{\AA}  contours; (d--f) HXR 10-20 keV
   and 25-45 keV maps (integrated over 12 seconds) over 1600~{\AA}
    ribbons; (g--i) NoRH 17 GHz I and V contours over HMI
   magnetograms (scaled between -2000 and 2000 G). See the electronic edition
of the Journal for a color version of this figure.
 \label{fig:maps}}
 \end{figure*}

Fig.~\ref{fig:maps}a--c shows the isocontours of EUV images taken by
SDO/AIA at 1600~{\AA} that reveal two highly sheared ribbons, almost
orthogonal to each other, instead of the more usual quasi-parallel
configuration.  The ribbons are located on opposite sides of the PIL
and extend along it.  Several bright kernels flash along the ribbons
during the impulsive phase.  The ribbons appear to be connected by the
coronal magnetic field, as seen in images at the higher-temperature
EUV wavelengths (Fig.~\ref{fig:maps}a--c).  The chromospheric ribbons
(1600~{\AA}) shift in position during the flare, as the hot coronal
loops (94~{\AA}) expand, forming post-flare hot loops which decrease
somewhat in apparent shear, though their final state is still highly
sheared. These flaring coronal loops connecting the two magnetic
domains around the flare can also be seen in the 171, 193, 211 and
335~{\AA} images.
%
%
\subsection{Hard X-rays and microwaves}
\subsubsection{Time evolution}
\begin{figure}
\plotone{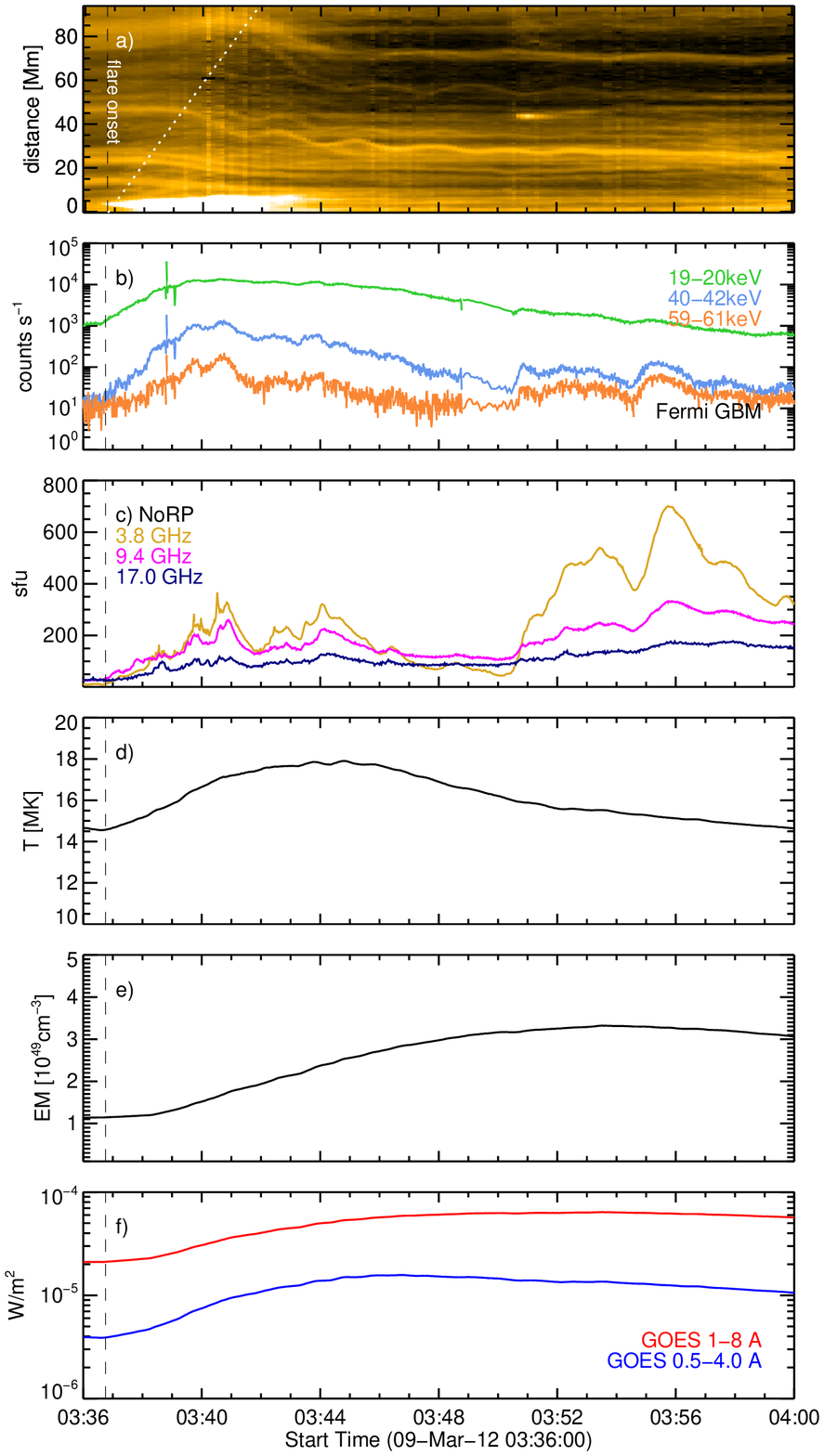}
\caption{Time evolution of the flare. a) time-position diagram of the
  slit; b) Fermi GBM HXR count-rate; c) NoRP lightcurves; d)
  temperature and e) emission measure ($EM$) derived from GOES data;
  f) GOES lightcurves.
The flare onset and the contraction onset of the four
  loops are indicated by the vertical dashed line and the dotted line
  respectively. See the electronic edition
of the Journal for a color version of this figure.
\label{fig:lightcurves}}
\end{figure}

We have shown the representative lightcurves for SXR, HXR, and
microwaves in Fig.~\ref{fig:lightcurves}, along with temperature $T$
and emission measure $EM$ derived from GOES data, and the EUV
time-position diagram from Fig.~\ref{fig:wiggles}. This also provides
a convenient reference for the different phases of the flare
development.  The high-energy HXR and microwave emissions of the main
flare started at 03:36:45~UT and peaked around 03:40:50~UT.  EUV
emission spatially integrated at wavelengths 171, 193, 211, 304, 1600
and 1700~{\AA} also peak around this time, with good correlation with
HXR and microwave main peaks, suggesting an association between the
non-thermal electrons, the ionisation equilibrium of coronal and
chromospheric EUV lines, and UV continuum emission in the
chromosphere. At 94, 131, and 335~{\AA} the emissions keep increasing,
peaking after the GOES SXR peak and with different delays: 03:59:15
(94~{\AA}), 03:53:11 (131~{\AA}), 04:02:05 (335~{\AA}), evidencing the
cooling of the coronal plasma.  Moreover, the peak time of
the impulsive phase (determined from HXR and microwave data) closely
matches the peak of the white-light flare signature reported by
\citet{2012A&A...544L..17H}, though their data only have one-minute
time cadence.

\subsubsection{Spatial characteristics}
\begin{figure*}
\plotone{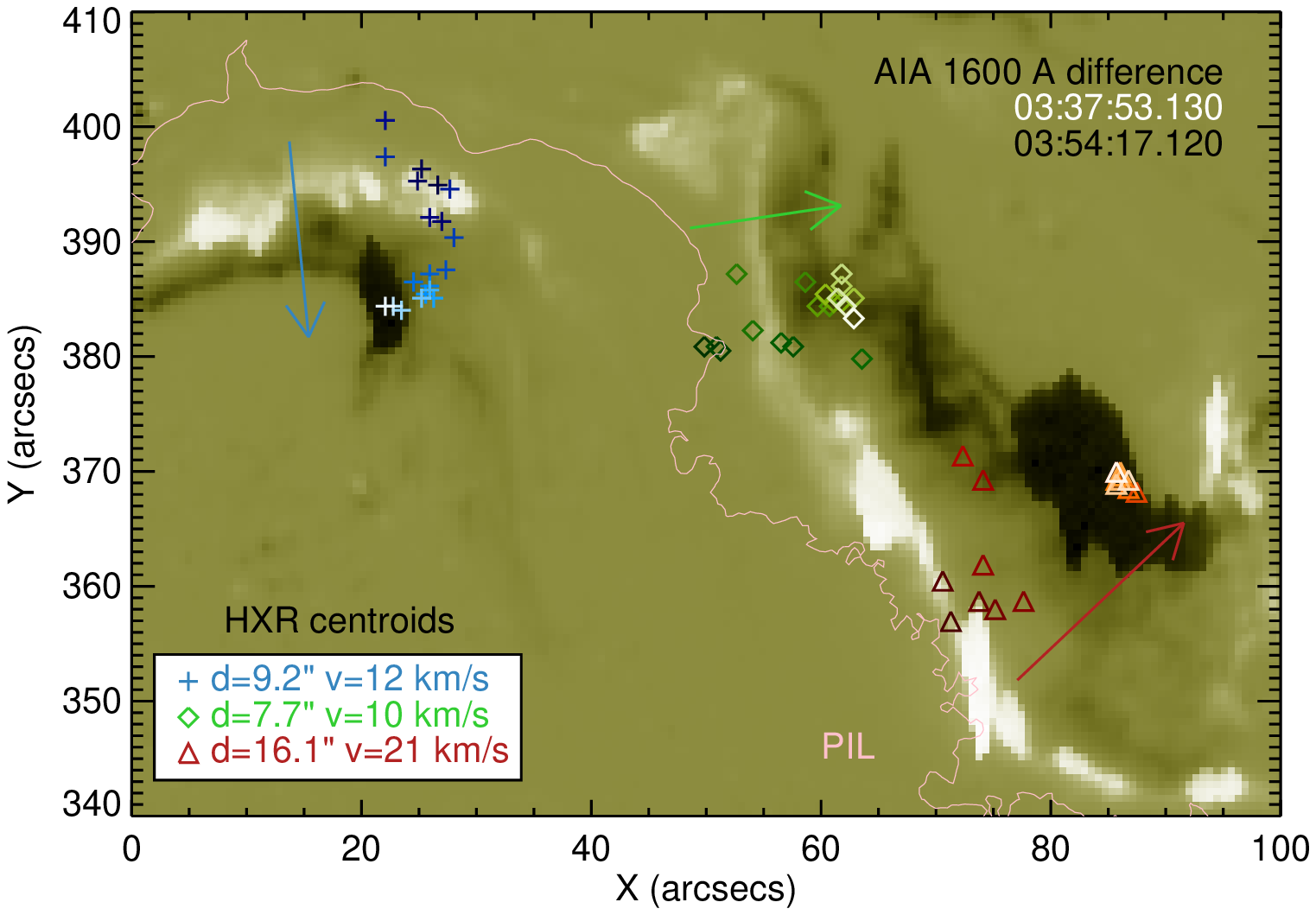}
 \caption{Centroid positions of the three main RHESSI HXR~25--45 keV
   sources (east, middle and west, indicated by crosses, diamonds and
   triangles, respectively) overlying the chromospheric ribbons seen by
   AIA 1600~{\AA}. The arrows show the direction of the sources'
   movements. The background image shows the difference 1600~{\AA} map,
   revealing the ribbons during the rise phase (03:37:53~UT) in white
   and during the late phase (03:54:17~UT) in black. See the electronic edition
of the Journal for a color version of this figure.
 \label{fig:hxrfp}}
 \end{figure*}

We constructed RHESSI CLEAN images at 10--20~keV and 25--45~keV at
selected intervals during the event.  The 25--45~keV images reveal the
presence of at least three distinct sources, which we identify as
chromospheric footpoints localized along the two UV ribbons.  They are
spatially coincident with the bright kernels seen in EUV images, as
shown in Fig.~\ref{fig:maps}d--f.  During the flare, there is at least
one footpoint source at each ribbon (at each side of the PIL), and
these sources move (Fig.~\ref{fig:hxrfp}) along with the
ribbons. The east footpoint (centroids indicated by crosses in
  Fig.~\ref{fig:hxrfp}) moves southwards, the middle footpoint
  (diamonds) moves westwards, while the west footpoint (triangles)
  moves northwest. The average distances and speeds are indicated in
  Fig.~\ref{fig:hxrfp}.  The 10--20~keV maps shown in
Fig.~\ref{fig:maps} are well associated with the coronal flaring loop
arcade seen in AIA images throughout the entire flare.
Microwave Stokes~I emission maps from NoRH at 17~and 34~GHz show an
extended source in excellent agreement with the two-ribbon structure
connected by low-lying coronal loops seen in EUV images
(Fig.~\ref{fig:maps}g--i).  Moreover, at least four distinct sources
can be identified at 17 and 34 GHz (indicated in
Fig.~\ref{fig:maps}h as I-IV), with two of them (II and III) in good
association with the HXR 25-45 keV footpoint sources, at each ribbon.
The implication is that during the impulsive phase the microwaves
could originate mainly at the footpoints, or near the ends of their coronal
loops, rather than from the main body of the loops
\citep[e.g.][]{2009ApJ...697..735R,2010ApJ...724..171R}. The rather
steep spectral index $\delta \approx 5.5$ derived from RHESSI HXR data
(and further confirmed by Fermi GBM data, see Table \ref{tab:hxr}),
indicates a lack of relativistic electrons, and thus strong magnetic
fields such as one might find at footpoints are required to produce
gyrosynchrotron emission at 17 and 34 GHz
\citep[e.g.][]{2006A&A...453..729S}.
\subsection{HXR, SXR and EUV pulsations}
\begin{figure}
   \plotone{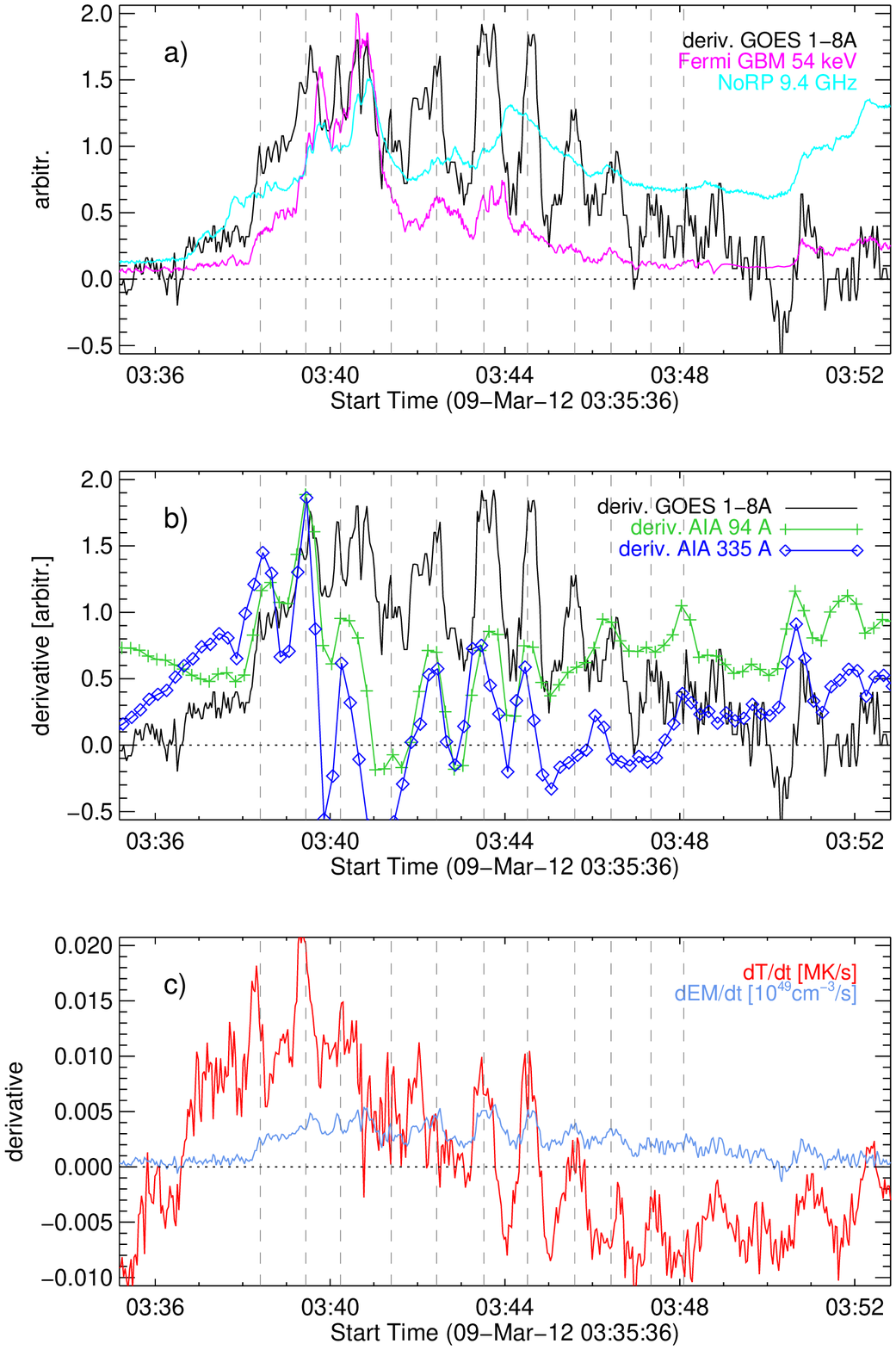}
   \caption{Oscillations of the flaring loop system, the vertical lines
     roughly indicate the ~60 second pulsations. a) normalised GOES
     time derivative, Fermi GBM count-rate at 54 keV, NoRP 9.4
     GHz; b) normalised time derivatives of the spatially integrated AIA
     maps over the flaring region at 94~{\AA} and 335~{\AA}; c)
     time derivatives of the temperature and emission measure from GOES. See the electronic edition
of the Journal for a color version of this figure.
 \label{fig:derivgoes}}
\end{figure}

The oscillations of the large-scale EUV loops during and after their
contraction is an obvious feature of this event, with oscillation
periods around 2--5 min, and a detailed analysis will be presented
elsewhere (Russell et al., in preparation).  There are also
shorter-period pulsations present, originating in hotter plasmas
closer to the flare core.  These are present in GOES data as well as
spatially-integrated AIA data.  The GOES time derivatives shown in
Fig.~\ref{fig:derivgoes}a reflect the Neupert behavior
\citep{1968ApJ...153L..59N} in this event, but also enhances the
visibility of oscillatory behavior at these soft X-ray temperatures.
These become apparent after the end of the HXR impulsive phase, at
03:42~UT, in the form of at least five quasi-periodic pulses about
1~min.  These quasi-periodic pulses are not so visible in the RHESSI HXR
data, but they are clearly seen in Fermi GBM lightcurves (see
Fig.~\ref{fig:derivgoes}a). We believe that this reflects the presence
of the RHESSI attenuators, automatically deployed in major flares.
This reduces the soft X-ray response relative to that of Fermi, which has
no such attenuators. Similar oscillations are observed in the
derivative lightcurves of 94 and 335~{\AA} AIA channels
(Fig.~\ref{fig:derivgoes}b) (the 131~{\AA} channel saturates during the
impulsive phase), and this mass of observational material makes it
clear that large portions of the entire active-region volume have been
set into persistent and semi-regular motions.

%
%
\section{Discussion}
\label{sec:discussion}

\subsection{Energetics of the event}
\label{sec:energy}
Using instantaneous values for the EM and T derived from GOES
(Fig.~\ref{fig:lightcurves}), we calculated the total thermal energy
content\footnote{All formulae are quoted in CGS units.} given by:
\begin{equation}
E_{th}=3k_BT\sqrt{EM \times V}\ ,
\label{eq:eth}
\end{equation}
as well as its derivative with respect to time, which is the rate of increase of the thermal energy. The radiative
(Eq. \ref{eq:lrad}) and conductive (Eq. \ref{eq:lcond}) loss rates are
\begin{equation}
L_{rad}=6\times 10^{-22} EM \left(\frac{T}{10^5}\right)^{1/2} 
\label{eq:lrad}
\end{equation}
and
\begin{equation}
L_{cond}=4\times 10^{-6} T^{7/2}L^{-1}A\ ,
\label{eq:lcond}
\end{equation}
using standard formulae \citep[e.g.][]{2009psf..book.....T}.

To estimate the volume in Eq.~\ref{eq:eth}, we obtained the coronal source area at
the 50\% level of the RHESSI 10-20 keV maps, in 46 intervals during
the main impulsive phase, and then took $V=A_{th}^{3/2}$.  The
average value $\langle V(t) \rangle$ was determined, giving $\langle
V(t) \rangle \geq V=1\times 10^{29}$~cm$^3$.  We use the radiative
loss rate calculation implemented in the SSW GOES routines with
parameters derived from the GOES observations. 
For the conductive loss rate, we have to estimate the
loop area and length. The area is obtained done by selecting footpoint
sources above 60\% of maximum (to resolve individual footpoints) on
the RHESSI 25--45~keV map at the peak at 03:40:06~UT. This gives $A
\approx 4.5 \times 10^{18}$ cm$^2$. The lower limit to the loop length
is $L\approx 8\times 10^9$cm, estimated from the AIA 94 {\AA} projected loop
lengths.
The final values that we arrive at are $E_{th} \ \approx 0.9 \times
10^{30}\ \mathrm{erg}$, $L_{rad} \ \approx 7 \times
10^{26}\ \mathrm{erg}~\mathrm{s}^{-1}$ and $L_{cond} \ \approx 4 \times
10^{28}\ \mathrm{erg}~\mathrm{s}^{-1}$, averaged over the peak
  (03:39:40--03:41:00~UT). During this interval the average net increase of
thermal energy is $\approx 1.6 \times 10^{28}$~erg~s$^{-1}$.  Using the value
of the (dominant) conductive energy loss rate implies that energy
input must be $\approx 6 \times 10^{28}$~erg~s$^{-1}$ during this peak.
 
Considering now the energy budget of the main burst, the thermal
energy from GOES before the flare (set by the smaller M-flares before
the main event) is $7 \times 10^{30}$ erg (at 03:38~UT) and $1.4
\times 10^{31}$ erg after the HXR peak (at 03:48~UT), giving a net
increase of $E_{net} \approx 7 \times 10^{30}$erg. So, integrating the
energy losses by radiation and conduction over the 10 minutes of the
main burst (03:38--03:48~UT), we have $5\times 10^{29}$ erg and
$3\times 10^{31}$ erg respectively, so $E_{loss} \approx 3\times
10^{31}$~erg. The value for the conductive loss rate is an upper limit
as it must decrease as the lower corona gets hotter. Now, simply
considering $E_{net}=E_{gain}-E_{loss}$, the energy input into the
thermal plasma during the 10 minute-period should be $E_{gain} \approx
3.7 \times 10^{31}$ erg. We have fitted the spatially integrated HXR
spectra observed by RHESSI to estimate the energy involved, assuming
the standard collisional thick-target model. We assume a low-energy
electron cut-off energy of 17~keV. Table~\ref{tab:hxr} summarizes the
results.  Using the RHESSI fitting results to determine the electron
energy rate, the energy of the non-thermal electrons (above 17~keV)
can be calculated, by integrating over this period, at
 $\approx 3.5 \times 10^{31}$~erg, which is consistent with  $E_{gain}$. 
\begin{table}
\caption{Hard X-ray summary (averaged over the peak
  03:39:40--03:41:00~UT)}
\begin{tabular}{ll}
\hline
Temperature & 21~MK \\
Emission measure & $0.6 \times 10^{49}$ cm$^{-3}$\\
Electron rate $> 17$~keV & $30 \times 10^{35}$ s$^{-1}$\\
Spectral index $\delta$ & 5.5\\
Electron energy & $8.5 \times 10^{30}$ erg \\
Peak electron energy rate & $1.1 \times 10^{29}$ erg~s$^{-1}$ \\
\hline
\label{tab:hxr}
\end{tabular}
\end{table}
\subsection{Oscillations in the flaring core}
The oscillations are seen in temperature and emission measure derived
from GOES data (Fig.~\ref{fig:derivgoes}c), and the estimates in
Sect. \ref{sec:energy} show that the radiative and conductive losses
are too slow to track them. Moreover, the time-coincidence of several
observables is intriguing: around 03:38~UT, at the onset of the
coronal implosion (see Fig.~\ref{fig:timing}), the AIA 94 and
335~{\AA} pulsations start (Fig.~\ref{fig:derivgoes}b) and occur
roughly in phase with the GOES oscillations, the GOES EM increases
rapidly and there is a peak in the rate of increase of the temperature
(Fig.~\ref{fig:derivgoes}c). We interpret these quasi-periodic changes
in temperature and emission measure as due to compressive oscillations
of the flaring loops, as equilibrium is restored following the
magnetic implosion. As kink-mode oscillations are only weakly
compressive
\citep{1983SoPh...88..179E,2005LRSP....2....3N,2008ApJ...676L..73V,2009A&A...503..213G},
most of the motion may be associated with global sausage modes or
longitudinal standing slow waves.  \citet{2011ApJ...740...90V}, using
{\em PROBA2/LYRA} data to study the flare SOL2010-02-08, found
oscillatory signatures at two distinct periods, $\sim 75$~s and
$\sim 8$~s, which were interpreted as standing (sausage) slow wave
and fast sausage mode, respectively. Following their interpretation,
we are tempted to favour the standing slow sausage mode. 
In any
case, we can derive a lower limit on the energy stored in these
standing waves from the variation of the thermal energy of the
confined plasma, noting that considerably more stored energy may be
present in the magnetic field itself. We consider that the average
energy associated with the oscillations is the root-mean-square of the
amplitude of the thermal energy derived from GOES (Eq.~\ref{eq:eth}),
after removing the slow-varying component.  For the 10-minute duration
of the main flare (03:38--03:48~UT) the average energy is $3 \times
10^{28}$ erg.  This is a first estimate of energy storage in the form
of large-scale standing waves in a flaring active region, and our
lower limit puts it at only a small fraction of the thermal energy
(Eq.~\ref{eq:eth}).

Compression of the loop magnetic field could be also associated with the
HXR pulsations \citep[e.~g.][]{1975ApJ...200..734B}, as a betatron acceleration mechanism can increase the
perpendicular momentum of electrons in the loop, via conservation of
the first adiabatic invariant, $p_\perp^2/B$, while the collapse of a
loop can further accelerate particles via the first-order Fermi
mechanism
\citep[e.g.][]{2005AstL...31..537B,2007AstL...33...54B}. According to
\citet{2007AstL...33...54B} starting from a hot thermal distribution
and accelerating by the betatron mechanism as the loop radius changes
will result in another thermal distribution; to obtain a non-thermal
distribution requires also loop collapse and Fermi acceleration. In
this event we have evidence for both. Exact expressions for the
spectral evolution are given in \citet{2007AstL...33...54B}, however
it was shown by \citet{2005AstL...31..537B} that the collapse of a
field of initial mirror ratio $b_m = B_\mathrm{max}/B_\mathrm{min}$
and temperature $T_0$ results in a non-thermal spectrum with
`effective temperature' $T_\mathrm{eff}$, characterizing the mean
kinetic energy of the trapped electrons, of $T_\mathrm{eff}=T_0(b_m
+2)/3$. To develop a value of $T_\mathrm{eff}$ = 17~keV (the
low-energy electron cut-off) from a thermal loop distribution of $T_0
\sim$ 1.3~keV (the typical thermal energy of the hot plasma) would
require $b_m \sim 37$, rather a high value. However, lower values of
$b_m$ could result in acceleration of part of the Maxwellian tail,
still giving significant non-thermal fluxes, while continuing
pulsations with a timescale not too different from the collision
timescale can lead to further acceleration even once the collapse has
stopped \citep{1997A&A...322..242K}. Alternatively, or more likely
acting along with the mechanisms above, the compression of the loop
would alter the magnetic trapping conditions by modulating the mirror
ratio $b_m$ and hence the precipitation rate of electrons into the
chromosphere \citep{1982SvAL....8..132Z}.
%
%
\section{Conclusions}
Many studies now, as cited above, have detected contracting motions in
the corona during solar flares that can be interpreted as the
implosion necessary to release energy.  Our results confirm this
behaviour in a particularly well-observed flare, SOL2012-03-09,
noting the close time association between the collapsing coronal loops
and impulsive phase of the flare.  We have added to the phenomenology
by noting the presence of oscillatory variations revealed by GOES soft
X-rays \citep[cf.][]{2012ApJ...749L..16D}, and have argued that these
result from eigenstates excited by the implosion and containing
compressive motions.  We speculate that the existence of such
persistent large-scale motions in the strong-field core of an active
region could play a role in medium-term (minutes) energy storage for a
solar flare. Our analysis can be summarised as follows:

1) The EUV observations shows at least four groups of coronal loops at
different heights overlying the flaring core undergoing fast
contraction during the impulsive phase of the flare. There are clear
variations of period with the location of the oscillatory loops within
the active region, with higher/longer loops having longer
periods. Also there is a clear outward motion through this structure
of an exciter with a projected velocity of $\sim$300~km~s$^{-1}$. This driver
is possibly associated with an EUV wave and/or a CME launch, both
identified in this event. After the passage of this driver a large
portion of the corona over the AR dims out, suggesting that the
coronal plasma was removed by the CME.  This sequence (implosion,
excitation wave, and oscillation) has been previously reported
\citep[e.g. ][]{2012ApJ...749...85G}, and we do not think that it is
uncommon
\citep{2009ApJ...696..121L,2009ApJ...706.1438J,2009ApJ...703L..23L,2010ApJ...714L..41L,2012ApJ...757..150L}. We
interpret the coronal implosion as a lack of magnetic support as
consequence of the energy release from the magnetic field
\citep[e.~g.][]{2000ApJ...531L..75H}. Such changes are entirely
consistent with the observed stepwise changes in the photospheric
magnetic field \citep{1994ApJ...424..436W,2005ApJ...635..647S}, with
the overall energy balance of the flaring active region, and with
the Alfv\'enic transport of energy in the impulsive phase
\citep{2008ApJ...675.1645F,2013ApJ...765...81R}.

2) The contraction of the magnetic structure starts around a minute
after the flare onset, and the rate of contraction is closely
associated with the intensity of the HXR and microwave emissions.
Highly-sheared chromospheric ribbons shift in position during the
flare, with coronal loops connecting the ribbons expanding and
reducing their shear state.  High-energy HXR footpoints are very well
associated with the ribbons, and follow their motions, and a
low-energy HXR (thermal) source is in excellent agreement with the
flaring coronal loops shown by EUV images. The HXR footpoint sources
(seen in this case also in microwave observations) reveal large
concentrations of energy release in the lower solar atmosphere, with
magnitude ($\approx 3.5\times 10^{31}$erg) consistent with the
thermal energy input ($E_{gain}\approx 3.7\times 10^{31}$erg) into the
flaring core after considering conduction ($L_{cond}\approx 3\times
10^{31}$erg) and radiative ($L_{rad}\approx 5\times 10^{29}$erg)
losses, to sustain the net thermal energy derived from observation
($E_{net}\approx 7\times 10^{30}$erg).  These observations are
comprehensive, and fully support the accepted hypothesis of
large-scale magnetic restructuring powering the several forms of
energy release. 

3) Pulsations in GOES and spatially integrated AIA data impose
changes in temperature and/or density in the flaring loops. We
interpret these pulsations basically as signatures of the complicated
compressive eigenstates of the active-region magnetic structure,
noting that such eigenstates may be excited even outside flare times
\citep{2013A&A...552A..57N}, and seen both in relatively hot or cold
coronal loop systems at thermal energies \citep{2012A&A...545A.129W}.
We have made a first crude estimate of the energy stored in the
eigenmodes based on the thermal energy of the hot plasma, but only as
a lower limit. We found an average energy of $\approx 3 \times
10^{28}$erg over the main flare (03:38--03:48~UT), less than 1\% of
the thermal energy. Moreover, such oscillations could change/modulate
the rate of acceleration, rate of precipitation and/or force new
acceleration directly
\citep[e.g.][]{2009SSRv..149..119N,2012SoPh..280..491K}, explaining
the HXR pulsations detected at roughly the same period as the SXR and
EUV pulsations. We speculate on whether these compressive oscillations could
be caused by the force balance after the fast magnetic
implosion. However, we do not perform a coronal
magnetoseismology analysis as it is beyond the scope of this work.

\acknowledgements
PJAS and LF acknowledge financial support by the European
Commission through HESPE (FP7-SPACE-2010-263086). LF acknowledges support from STFC grant ST/l001808.
AJBR was a research fellow of the Royal Commission for
the Exhibition of 1851 at the University of Glasgow. 
HSH acknowledges support from NASA under contract NNX11AP05G for RHESSI.

Facilties: \facility{SDO(AIA)}, \facility{SDO(HMI)},
\facility{RHESSI}, \facility{NoRH}, \facility{NoRP}, \facility{GOES},
\facility{Fermi(GBM)}

\bibliographystyle{plainnat}
\bibliography{refs_rhessi}
\end{document}